\documentclass[journal]{IEEEtran}

\usepackage{hyperref}
\usepackage{graphicx}
\graphicspath{ {./} }

\begin{document}
\title{Before and After: \\ Machine learning for perioperative patient care}

\author{%
        {Iuliia Ganskaia, \textit{Independent Researcher}}, {Stanislav Abaimov, \textit{University of Rome, Tor Vergata}}

}

\markboth{Submitted for peer-review to IEEE Access 2021}%
{Ganskaia \MakeLowercase{\textit{et al.}}: Bare Demo of IEEEtran.cls for IEEE Journals}
\maketitle

\begin{abstract}
    For centuries nursing has been known as a job that requires complex manual operations, that cannot be automated or replaced by any machinery. All the devices and techniques have been invented only to support, but never fully replace, a person with knowledge and expert intuition. With the rise of Artificial Intelligence and continuously increasing digital data flow in healthcare, new tools have arrived to improve patient care and reduce the labour-intensive work conditions of a nurse.
    
    This cross-disciplinary review aims to build a bridge over the gap between computer science and nursing. It outlines and classifies the methods for machine learning and data processing in patient care before and after the operation. It comprises of Process-, Patient-, Operator-, Feedback-, and Technology-centric classifications. The presented classifications are based on the technical aspects of patient case. 

\end{abstract}

\IEEEpeerreviewmaketitle

\section{Introduction}
    Artificial Intelligence (AI) has already greatly contributed to the development of humankind. AI is applied in research, economic and strategic planning, warfare, entertainment, and in numerous branches of healthcare. Various awards and funding opportunities across the globe are announced to boost the growth of AI for health, care, and well-being\footnote{\href{https://www.england.nhs.uk/aac/what-we-do/how-can-the-aac-help-me/ai-award/}{Artificial Intelligence Award in Health and Care Award, NHS}}.

    \textit{“Progress in robotics and artificial intelligence should be oriented "towards respecting the dignity of the person and of Creation,”} - Pope Francis\footnote{\href{https://www.vaticannews.va/en/pope/news/2020-11/pope-francis-november-prayer-intention-robotics-ai-human.html}{"Pope’s November prayer intention: that progress in robotics and AI “be human”", Vatican News, November 2020}}.
        
    Machine learning has already been extensively researched for medical and pharmacological purposes. Thus, following Maslow's hierarchy of needs, this review presents the subject of machine-learning-powered technologies and methods directly related to patient care before and after the operation. Preoperative diagnostics and real-life monitoring during the operation are the essential steps in emergency response, patient treatment. However, postoperative recovery of the patient takes significantly longer and can be interrupted by even the smallest changes in the patient's condition and behaviour.
    
    Machine learning can be applied to save lives directly and indirectly. A great number of research papers has already been published about the application of machine learning for diagnostics and innovative drug discovery. However, emerging technologies in patient care still remain majorly unexplored. 
    
    This review proposes a set of classifications at the technical level of devices and data processing, that is specific to medical and patient care equipment, as well as health-related personal data. In addition, the review lists current and future challenges for the field of study. The review shows, that there are numerous methods, in which machine learning can improve the effectiveness of medical care, speed up the recovery process, and reduce stress in patients.

\section{Background}
    
    The development of the new applications requires sufficient knowledge in both patient care and computer sciences, as patient care has a set of unique characteristics of human-machine interaction, that are not present in other fields, such as engineering, entertainment, and marketing.
    
    \subsection{Patient care}
        In emergency situation, monitoring vital conditions and symptoms before the operation, is essential in order to determine the correct actions and administered medications. Pre-existing medical conditions and stress can affect the behaviour and reaction to medications, anesthetics, and provided aid in general.
        
        Monitoring and support after the operation is equally important, to ensure stable and rapid recovery of the patient.
        
        Some of the electronic and connected IoT devices, used for patient care include: computers, tablets, smart beds, smart life support, smart mobility vehicles, monitoring devices and sensors, experimental robotic systems \cite{locsin2018can}.

    \subsection{Maslow's hierarchy of needs}

\begin{figure}[h!]
            \includegraphics[width=8cm]{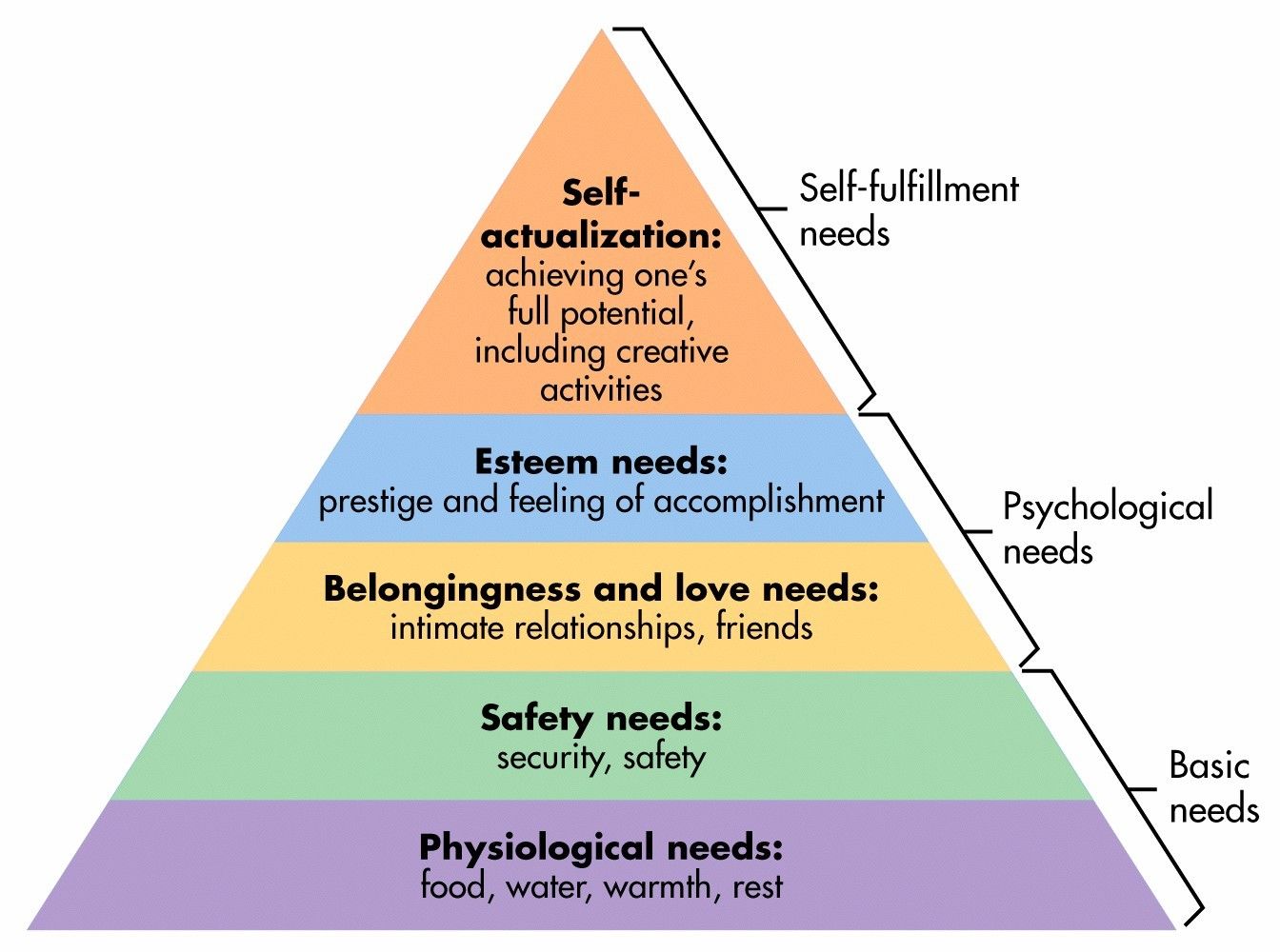}
            \caption{Maslow's hierarchy of needs}
            \label{fig:maslow}
        \end{figure}

        Published in 1943 \cite{maslow1943theory}, the theory of human motivation and the hierarchy of needs\footnote{\href{https://sites.psu.edu/rclerin/2015/04/10/hierarchy-of-needs/}{Image from Pennsylvania State University, 2015}} (see Fig. \ref{fig:maslow}) are still used in patient care practice. Each person's needs must be met individually in order for them to feel satisfied, cared for and cooperative:
        \begin{itemize}
            \item Identify the level of care required for the patient. 
            \item Evaluate patient communication to determine what needs are not being met. 
            \item Adjust your behavior towards your patients to show acknowledgement of those needs. 
            \item Allow yourself to form a bond with patients in extended care. 
        \end{itemize}

    \subsection{Machine Learning}
        Machine learning (ML) is a set of methods and algorithms, that allows a computerised device to learn from already existing data.

        ML can be supervised, unsupervised, semi-supervised, reinforcement learning. Any type of ML requires data in a form of a data set (N-dimensional matrix or a database) or a continuous stream of data. The preprocessed data set is used for the training of the model, and then the trained model is used to identify or "predict" the most likely outcomes of the new data (typically not present in the training data set).

\section{Existing applications and research}\label{sec:applications}
    Patient care has always been a task, too complex for any machinery to be trusted with. To ensure the gradual and successful recovery of the patients, a nurse needs knowledge, flexible decision making, and a degree of expert intuition.
    
    Earlier reviews addressed the issues of physical assistance in nursing using robotics systems \cite{locsin2018can}. Robert \cite{robert2019artificial} presents one of the first non-technical overviews of potential AI applications for nursing. A 2021 review\cite{buchanan2020nursing} presented the way to structure the data for AI-based decision support systems. Another 2021 survey\cite{ronquillo2021artificial} provides recommendations for Education, Practice, Research, and Leadership, as well as a top-level review of ethical and legal issues, and future implications of the use of AI for nursing.  
    
    \subsection{Smart devices in patient care}
        
        The market of Internet-of-Things and connected devices is continuously growing and it will keep growing over the upcoming decades. Many of those devices have been programmed and named "smart devices". There are several distinct types of smart devices that are directly relevant for patient care and can be augmented with machine learning.
        
        \textbf{Smart beds and smart medical beds}
            
            An average person with healthy sleeping habits spends approximately a third of life sleeping. Post-operative patients spend most of their time in bed to maximise recovery.
            
            Machine learning is used for the improvement of consumer beds, directly linking electronics to rest and recovery in humans. Similarly, ML is used for smart medical beds to maximise patient recovery and monitoring capacity \cite{ghersi2018smart}.
            
        \textbf{Emergency response}
            
            Machine learning can provide early diagnostics and guidance, playing an advisory role in emergency response, such as dynamic detection of cerebral ischemia\cite{Megjhani2020.ischemia} or discharge prediction with spinal cord injury\cite{Fan2021.06.26.discharge}.
            
        \textbf{Ventilation control}
        
            ML is applied for pressure-controlled ventilation \cite{machinelearningventilation2021} for patients with breathing conditions, especially crucial for the detection of anomalies and new critical viral infection or kinetic damage of the tissues.
            
        \textbf{Smart mobility vehicles}
            
            A wide range of mobility devices can be used to the patients with limited abilities during recovery. Those vehicles incluse autonomous power-operated scooters, as well as limb, back, or full body exoskeletons \cite{Zelik2021.07.22.exo}. Robotic and advanced bionic prosthetic limbs use machine learning for muscle and neural signal processing to return or even enhance mobility to patients with disabilities.
            
        \textbf{Remote patient monitoring}

            Remote patient monitoring is the most common application of IoT devices for healthcare. IoT devices can automatically collect health metrics like heart rate, blood pressure, temperature, and more from patients who are not physically present in a healthcare facility, eliminating the need for patients to travel to the providers, or for patients to collect it themselves.

            When an IoT device collects patient data, it forwards the data to a software application where healthcare professionals and/or patients can view it. Algorithms may be used to analyze the data in order to recommend treatments or generate alerts. For example, an IoT sensor that detects a patient’s unusually low heart rate may generate an alert so that healthcare professionals can intervene.

            A major challenge with remote patient monitoring devices is ensuring that the highly personal data that these IoT devices collect is secure and private.
            
            In addition to general health data, wearable devices can passively\cite{Gadaleta2021.wearable.covid} detect the symptoms of viral infections, such as COVID-19 that can be used for early warming and epidemic management.

        \textbf{Glucose monitoring}

            Glucose levels impact the behaviour and well-being of the patient before and after the operation. Glucose levels can fluctuate and cannot be monitored continuously with conventional manual methods, even with periodic testing. Smart IoT devices can address these challenges by providing continuous, automatic monitoring of glucose levels in patients. Furthermore, such devices eliminate the need to keep records manually, and they can alert patients when glucose levels are problematic.
        
        \textbf{Heart-rate monitoring}

            Like glucose, monitoring heart rates can be challenging, even for patients who are present in healthcare facilities. Periodic heart rate checks do not guard against rapid fluctuations in heart rates, and conventional devices for continuous cardiac monitoring used in hospitals require patients to be attached to wired machines constantly, impairing their mobility.
            
            For the past decade, a variety of small IoT devices have become commercially available in an affordable way for heart rate monitoring, freeing patients to move around as they like while ensuring that their hearts are monitored continuously. Guaranteeing ultra-accurate results remains a challenge, but most modern devices can deliver accuracy rates of about 90 percent or better and provide sufficient early warming in case of rapid raise of heart rate.
            
        \textbf{Hand hygiene monitoring}

            Traditionally, there has never been a verifiable way to ensure that providers and patients inside a healthcare facility washed their hands properly in order to minimize the risk of spreading contagion.

            With the COVID-19 popularisation of regular hand sensitisation the issue has become less severe. Furthermore, many hospitals and other health care operations use IoT devices to alert people to sanitise their hands when they enter hospital rooms. The devices can even give instructions on how best to sanitise to mitigate a particular risk for a particular patient. These devices can reduce infection rates by more than 60 percent in hospitals.
            
        


            
        
        \textbf{Connected Inhalers}
        
            Conditions such as asthma or COPD often involve attacks that come on suddenly, with little warning. IoT-connected inhalers can help patients by monitoring the frequency of attacks, as well as collecting data from the environment to help healthcare providers understand what triggered an attack.

            In addition, connected inhalers can alert patients when they leave inhalers at home, placing them at risk of suffering an attack without their inhaler present, or when they use the inhaler improperly.
        
        \textbf{Ingestible sensors}
        
            Collecting data from inside the human body is typically a messy and highly disruptive affair. No no enjoys having a camera or probe stuck into their digestive tract, for example.

            With ingestible sensors, it is possible to collect information from digestive and other systems in a much less invasive way. They provide insights into stomach PH levels, for instance, or help pinpoint the source of internal bleeding.

            These devices must be small enough to be swallowed easily. They must also be able to dissolve or pass through the human body cleanly on their own. Several companies are hard at work on ingestible sensors that meet these criteria.
        
        \textbf{Behaviour advisory systems}
            
            It is impossible to predict behaviour of a healthy human. It is even harder to predict the actions of a distressed patient in emergency condition. Modern technologies can provide smart solutions, that analyse the changes in the patients condition and yield adaptive guidance and reminders in activities, such as rest, exercises, hydration, and nutrition.
        
        \textbf{Connected glasses and contact lenses}
        
            In the near future ocular devices can collect additional . Smart contact lenses provide another opportunity for collecting healthcare data in a passive, non-intrusive way. They could also, incidentally, include microcameras that allow wearers effectively to take pictures with their eyes, which is probably why companies like Google have patented connected contact lenses.

            Whether they are used to improve health outcomes or for other purposes, smart lenses promise to turn human eyes into a powerful tool for digital interactions.
            
        \textbf{Public health}
        
            Smart cities have the capacity to monitor health of the general population, forecast and combat the spread of infectious diseases \cite{Ezugwu2020.Smartcity}. Using health data it is possible to predict emergency admission in specific parts of a city or a country \cite{Liley2021.scotland}.
        
        \textbf{Support for patients with disabilities}
        
            The care for patients with disabilities, who can have impaired speech, vision, or hearing, can be improved by the use of devices and sensors, that provide interpretation and additional information about the sensations and specific requests.
        
        \textbf{Training and Evaluation of preparedness}
        
            A Data-Driven approach using machine learning can be used to improve training programs for nurses and for further evaluation of acquired skills \cite{Liu2020.training, education_ai}.
            
    \subsection{Relevant applications of medical AI}
        Beyond patient care, some medical applications of the ML can be directly related to patient's successful recovery.
    
        \textbf{Drug discovery and toxicology}
    
            A large percentage of candidate drugs fail to win regulatory approval. These failures are caused by insufficient efficacy (on-target effect), undesired interactions (off-target effects), or unanticipated toxic effects. Research has explored use of deep learning to predict the biomolecular targets, off-targets, and toxic effects of environmental chemicals in nutrients, household products and drugs.
            
            AtomNet is a deep learning system for structure-based rational drug design\cite{wallach2015atomnet} AtomNet was used to predict novel candidate biomolecules for disease targets such as the Ebola virus and multiple sclerosis.
            
            In 2017 graph neural networks were used for the first time to predict various properties of molecules in a large toxicology data set. In 2019, generative neural networks were used to produce molecules that were validated experimentally all the way into mice.
        
        \textbf{Medical Image Analysis}

            Deep learning has been shown to produce competitive results in medical application such as cancer cell classification, lesion detection, organ segmentation and image enhancement \cite{Litjens_2017, 8265228}.
        
        

\section{Classification}

    The applications, presented in Section~\ref{sec:applications}, can be classified by the level of interaction with the care process, with the patient, with the operator (i.e., authorised care staff), the level of intensity (i.e., from low to high), and the volume of technological requirements and data volume.

    \subsection{Process-centric}
        Process-centric approach aims to group the devices by the frequency of data collection, volume of data, direct or indirect interaction with the patient, and speed of making decisions.
        \begin{itemize}
            \item Primary - the device collects data in real time and outputs the current state of the patient.
            \item Secondary - the device collects data over time to create a recovery profile, and make forecasts on the potential changes and recovery speed.
            \item Statistical - data is collected over time from multiple patients, analysed and further used to advise care staff on similar patients and deviations of the specific patient from previous similar cases.
        \end{itemize}
        
    \subsection{Patient-centric}
        Patient-centric classification is based on the actions, required from the patient in order to engage with the device.
        \begin{itemize}
            \item none: used by statistical methods
            \item passive: wear and/or regularly charge the device
            \item active: perform specific actions while the device is actively collecting data
        \end{itemize}

    \subsection{Operator-centric}
        Operator-centric classification is based on the actions, required from the nurses or other care staff:
        \begin{itemize}
            \item none
            \item passive: the device needs to be only activated or placed in the proximity or in contact with the patient
            \item active: the device needs to be activated, and control actions are needed to ensure the adequate operation of the device.
        \end{itemize}

    \subsection{Feedback-centric}
        Feedback-centric classification outlines the way that the device interacts with the patient:
        \begin{itemize}
            \item no-feedback: the device is placed close to or on the patient and performs data collection
            \item feedback: the device sends feedback signals to the patient, either visual, auditary or tactile. 
            \item data: output data is displayed on the screen of the device and/or parsed to an application or internal webserver.
        \end{itemize}

    \subsection{Technology-centric}
        
        Technology-centric classification is based on the technical aspects of device manifacturing and data processing, and divided into three sub-categories. Size or scale of the device, volume and origin of data, and level of provided privacy.
    
        Size and computational power of the electronic device:
        \begin{itemize}
            \item small - a small device, e.g., with sensors
            \item medium - a device that can be placed at the bedside of the patient
            \item large - a large device that requires a patient to be placed inside the device
            \item server - computer or a physical or virtual network of computers, that is used for data mining and processing
        \end{itemize}

        Volume and origin of data:
        \begin{itemize}
            \item Small volume, real-time - e.g., 1-5 variable parameters to monitor the current state of the recovering post-operative patient
            \item Large volume, real-time - e.g., full-body analysis for pre-operational decision making
            \item Small volume, statistical - e.g., heart-rate, blood pressure, sugar levels, oxygen levels in blood, to observe changes over time
            \item Large volume, statistical - e.g., regular full-body analysis over and extended period of time.
        \end{itemize}

        With the constantly growing number of interconnected devices and growing cyber crime, the concerns of personal privacy are rising every year. Thus, the data can be classified by the level of detail it provides about the patient:
        \begin{itemize}
            \item anonymous data - cannot be attributed back to the patient, without prior knowledge
            \item confidential data - can be attributed back to the patient, but is stored securely and cannot be accessed without authorisation.
            \item public records - publicly available data
        \end{itemize}

\section{Future challenges in patient care}

    The growing importance of machine learning in medical systems cannot be overstated. By 2025, research predicts that global AI healthcare spending will equal \$36.1 billion\footnote{\href{https://healthitanalytics.com/news/artificial-intelligence-in-healthcare-spending-to-hit-36b}{"Artificial Intelligence in Healthcare Spending to Hit \$36B", December 2018}}.
    
    "The strategy comes in the week new data shows global investors poured £13.5 billion into more than 1,400 UK private technology firms between January and June [2021] - more than that achieved in other large tech markets Germany, France and Israel combined." \footnote{\href{https://www.gov.uk/government/news/new-ten-year-plan-to-make-britain-a-global-ai-superpower}{"New ten-year plan to make the UK a global AI superpower", GOV.UK, 22 Sept 2021}}

    \subsection{Post-operative care and general well-being}
    
        Post-operative care is more engaged than the general monitoring of well-being after the "check out". However, in both cases machine learning aims to achieve the same goal - steady recovery of the patient and general well-being.
    
    \subsection{Current applications of machine learning}
    
        The vast majority of the published methods do not yet have the capacity for real-time  monitoring of vital data, to ensure the successful advisory role of ML-based software. Machine learning currently is only used for the optimisation of data collection and analysis for further linear decision making.
       
    \subsection{Trust in new technology}
    
        The notion of trust in machine learning. Medical staff requires additional training. Modern machine learning applications have not reached human level of decision-making accuracy. However, ML allows to analyse large volumes of data a lot quicker than a human operator, and provide the most likely outcome based on the previously collected data.

    \subsection{Cyber security}

        With the use of constantly advising Information Technologies, the concerns about privacy and security equally grow.

        Death, caused by a cyber attack on healthcare systems was registered in September 2020 \footnote{\href{https://www.nytimes.com/2020/09/18/world/europe/cyber-attack-germany-ransomeware-death.html}{Cyber Attack Suspected in German Woman’s Death, The New York Times, 2020}}. Another death, caused by a cyber attack, was registered in 2021\footnote{\href{https://www.pandasecurity.com/en/mediacenter/security/first-ransomware-death/}{Is this the first ransomware death in the USA?, Panda Security, 2021}}, and was proven in court, creating a precedent.
     
    \subsection{Ethical concerns}
    
        Similar to Lethal Autonomous Weapon Systems, medical system, powered by machine learning, cause ethical concerns, as they are responsible for human lives. Global experts have already voiced their concerns in open letters regarding Autonomous Weapon Systems, advising international organisation and national governments to ban the use of lethal systems. However, medical robotic devices can save lives, instead of taking them.
        
    \subsection{Standardisation and Regulation}    
        
        "The UK National Screening Committee (UK NSC) has approved interim guidance for those proposing to use AI in breast screening mammography. We will also shortly produce general guidance about when and how companies should engage with PHE about their AI products." \footnote{\href{https://phescreening.blog.gov.uk/2019/03/14/new-guidance-for-ai-in-screening/}{New guidance for AI in screening, Public Health England, 2019}}
    
    \subsection{Future applications of machine learning}
    
        Entrusting decision-making to autonomous drug delivery systems can cause ethical concerns, that need to be addressed through the development of trusted technologies and legal standards.

    \subsection{Ageing technology and cyber security}

        Similarly to the industrial control systems, medical facilities can accumulate a combination of outdated and modern technology. In medicine the hardware and software is directly or indirectly responsible for human lives, the fast-growing market of smart and IoT devices with potential cyber vulnerabilities results in high health risks.

\section{Conclusion}
    The article presents the review of the applications of Artificial Intelligence and Machine Learning for patient care. This includes real-time monitoring, reporting, diagnostics, and decision-making for supervised and unsupervised robotic first aid.
    
    With the growing population and the increasing complexity of health knowledge, the world is in desperate need of new tools and methods of healthcare. Machine learning can and will improve the quality of patient care.
    
    Machine learning can improve the effectiveness of medical care, speed up the recovery process, and reduce stress in patients. However, machine learning is a promising technology, which is not sufficiently developed to be fully entrusted with human lives, and can be only used as an experimental advisory approach.



\ifCLASSOPTIONcaptionsoff
  \newpage
\fi

\bibliographystyle{plain} 

\bibliography{main}

\begin{thebibliography}{10}

\bibitem{buchanan2020nursing}
Christine Buchanan, M~Lyndsay Howitt, Rita Wilson, Richard~G Booth, Tracie
  Risling, and Megan Bamford.
\newblock Nursing in the age of artificial intelligence: protocol for a scoping
  review.
\newblock {\em JMIR research protocols}, 9(4):e17490, 2020.

\bibitem{education_ai}
Christine Buchanan, M~Lyndsay Howitt, Rita Wilson, Richard~G Booth, Tracie
  Risling, and Megan Bamford.
\newblock Predicted influences of artificial intelligence on nursing education:
  Scoping review.
\newblock {\em JMIR Nursing}, 4(1):e23933, Jan 2021.

\bibitem{Ezugwu2020.Smartcity}
Absalom~E. Ezugwu, Ibrahim~A.T. Hashem, Olaide~N. Oyelade, Haruna Chiroma,
  Mohammed~A. Al-Garadi, Idris~N. Abdullahi, Olumuyiwa Otegbeye, Amit~K.
  Shukla, and Mubarak Almutari.
\newblock A novel smart city based framework on perspectives for application of
  machine learning in combatting covid-19.
\newblock {\em medRxiv}, 2021.

\bibitem{Fan2021.06.26.discharge}
Guoxin Fan, Huaqing Liu, Sheng Yang, Libo Luo, Lunji Wang, Mao Pang, Bin Liu,
  Liangming Zhang, Lanqing Han, and Limin Rong.
\newblock Discharge prediction of critical patients with spinal cord injury: a
  machine learning study with 1485 cases.
\newblock {\em medRxiv}, 2021.

\bibitem{8265228}
Gustav Forslid, Håkan Wieslander, Ewert Bengtsson, Carolina Wählby,
  Jan-Michael Hirsch, Christina~Runow Stark, and Sajith~Kecheril Sadanandan.
\newblock Deep convolutional neural networks for detecting cellular changes due
  to malignancy.
\newblock In {\em 2017 IEEE International Conference on Computer Vision
  Workshops (ICCVW)}, pages 82--89, 2017.

\bibitem{Gadaleta2021.wearable.covid}
Matteo Gadaleta, Jennifer~M. Radin, Katie Baca-Motes, Edward Ramos, Vik
  Kheterpal, Eric~J. Topol, Steven~R. Steinhubl, and Giorgio Quer.
\newblock Passive detection of covid-19 with wearable sensors and explainable
  machine learning algorithms.
\newblock {\em medRxiv}, 2021.

\bibitem{ghersi2018smart}
Ignacio Ghersi, Mario Mari{\~n}o, and M{\'o}nica~Teresita Miralles.
\newblock Smart medical beds in patient-care environments of the twenty-first
  century: a state-of-art survey.
\newblock {\em BMC medical informatics and decision making}, 18(1):1--12, 2018.

\bibitem{Liley2021.scotland}
James Liley, Gergo Bohner, Samuel~R. Emerson, Bilal~A. Mateen, Katie Borland,
  David Carr, Scott Heald, Samuel~D. Oduro, Jill Ireland, Keith Moffat, Rachel
  Porteous, Stephen Riddell, Nathan Cunningham, Chris Holmes, Katrina Payne,
  Sebastian~J. Vollmer, Catalina~A. Vallejos, and Louis J.~M. Aslett.
\newblock Development and assessment of a machine learning tool for predicting
  emergency admission in scotland.
\newblock {\em medRxiv}, 2021.

\bibitem{Litjens_2017}
Geert Litjens, Thijs Kooi, Babak~Ehteshami Bejnordi, Arnaud Arindra~Adiyoso
  Setio, Francesco Ciompi, Mohsen Ghafoorian, Jeroen~A.W.M. van~der Laak, Bram
  van Ginneken, and Clara~I. Sánchez.
\newblock A survey on deep learning in medical image analysis.
\newblock {\em Medical Image Analysis}, 42:60–88, Dec 2017.

\bibitem{Liu2020.training}
Yanmei Liu and Yuwen Chen.
\newblock A data-driven evaluation approach for assessing student nurse
  training effectiveness in clinical practice using a fuzzy mathematics model.
\newblock {\em medRxiv}, 2020.

\bibitem{locsin2018can}
Rozzano~C Locsin and Hirokazu Ito.
\newblock Can humanoid nurse robots replace human nurses.
\newblock {\em Journal of Nursing}, 5(1):1--6, 2018.

\bibitem{maslow1943theory}
Abraham~Harold Maslow.
\newblock A theory of human motivation.
\newblock {\em Psychological review}, 50(4):370, 1943.

\bibitem{Megjhani2020.ischemia}
Murad Megjhani, Kalijah Terilli, Ayham Alkhachroum, David~J. Roh, Sachin
  Agarwal, E.~Sander Connolly, Angela Velazquez, Amelia Boehme, Jan Claassen,
  and Soojin Park.
\newblock Dynamic detection of delayed cerebral ischemia using machine
  learning.
\newblock {\em medRxiv}, 2020.

\bibitem{robert2019artificial}
Nancy Robert.
\newblock How artificial intelligence is changing nursing.
\newblock {\em Nursing management}, 50(9):30, 2019.

\bibitem{ronquillo2021artificial}
Charlene~Esteban Ronquillo, Laura-Maria Peltonen, Lisiane Pruinelli, Charlene~H
  Chu, Suzanne Bakken, Ana Beduschi, Kenrick Cato, Nicholas Hardiker, Alain
  Junger, Martin Michalowski, et~al.
\newblock Artificial intelligence in nursing: Priorities and opportunities from
  an international invitational think-tank of the nursing and artificial
  intelligence leadership collaborative.
\newblock {\em Journal of advanced nursing}, 2021.

\bibitem{machinelearningventilation2021}
Daniel Suo, Cyril Zhang, Paula Gradu, Udaya Ghai, Xinyi Chen, Edgar Minasyan,
  Naman Agarwal, Karan Singh, Julienne LaChance, Tom Zajdel, Manuel Schottdorf,
  Daniel Cohen, and Elad Hazan.
\newblock Machine learning for mechanical ventilation control.
\newblock {\em medRxiv}, 2021.

\bibitem{wallach2015atomnet}
Izhar Wallach, Michael Dzamba, and Abraham Heifets.
\newblock Atomnet: A deep convolutional neural network for bioactivity
  prediction in structure-based drug discovery, 2015.

\bibitem{Zelik2021.07.22.exo}
Karl~E. Zelik, Cameron~A. Nurse, Mark~C. Schall, Richard~F. Sesek, Matthew~C.
  Marino, and Sean Gallagher.
\newblock Assessing the effect of back exoskeletons on injury risk during
  material handling.
\newblock {\em medRxiv}, 2021.

\end{thebibliography}


\end{document}